\documentclass[proof]{WileyASNA-v1}
\usepackage{graphicx,epstopdf}
\usepackage{amsmath,mathtools}
\usepackage{amssymb}

\def\degr{\hbox{$^\circ$}}

\articletype{Original Article}%

\received{20 March 2020}
\revised{2020}
\accepted{}

\begin{document}

\title{PTF1J2224$+$17: a short-period, high-field polar}

\author[1]{A. D. Schwope*}

\author[2]{B. D. Thinius}

\authormark{Schwope \& Thinius}

\address[1]{\orgname{Leibniz-Institute for Astrophysics Potsdam (AIP)}, \orgaddress{\state{Potsdam}, \country{Germany}}}

\address[2]{\orgname{Inastars Observatory}, \orgaddress{\state{Potsdam}, \country{Germany}}}

\corres{*A.D. Schwope, Leibniz-Institute for Astrophysics Potsdam (AIP), An der Sternwarte 16, Potsdam, Germany. \email{aschwope@aip.de}}

\abstract{We present time-resolved photometry of the cataclysmic variable (CV) PTF1J2224$+$17 obtained during 4 nights in October 2018 and January 2019 from Inastars observatory. The object is variable on a period of 103.82\,min.
Archival Catalina Real-Time Transient Survey (CRTS), PTF, and ZTF-data show frequent changes between high and low states.
Based on its photometric properties and the cyclotron humps in the identification spectrum 
the object is certainly classified as an AM Herculis star (or polar) with a likely magnetic field strength of $B\sim65$\,MG. Its accretion duty cycle was estimated from nine years of photometric monitoring to be about 35\%. }


\maketitle

\section{Introduction}\label{sec1}
Cataclysmic variables (CVs) are close mass-transferring binaries with a late-type donor star and a white dwarf accretor. Depending on the magnetic field strength of the white dwarf, an accretion disk forms or not \citep[for comprehensive descriptions of the properties of magnetic and non-magnetic CVs see][]{warner95}. The disks in short-period non-magnetic CVs become unstable from time to time and show dwarf nova outbursts with amplitudes between $2-8$ mag. Such objects are uncovered in considerable number in contemporaneous variability surveys like the CRTS \citep{drake+14}. 

In magnetic CVs accretion happens quasi-radially with out a disk as intermediary. This leads to strong X-ray emission and pronounced variability 
on the spin period of the white dwarf. In CVs with the strongest magnetic fields, the AM Herculis stars or polars, the spin and the orbit are synchronized so that only one kind of periodicity can typically be recognized. The optical phenomenology of such objects might be complex. Orbital (spin) phase modulation in the optical regime is due to fore-shortening, due to very pronounced cyclotron beaming, irradiation, eclipses and self-occultations. On top of it  changes between states of high and low accretion may occur on unpredictable time scales and for unpredictable time intervals. Eclipses, cyclotron beaming and accretion-rate changes may also lead to brightness changes of up to 6 mag. 

\begin{figure}[t]
\centerline{\includegraphics[width=0.5\textwidth,clip=]{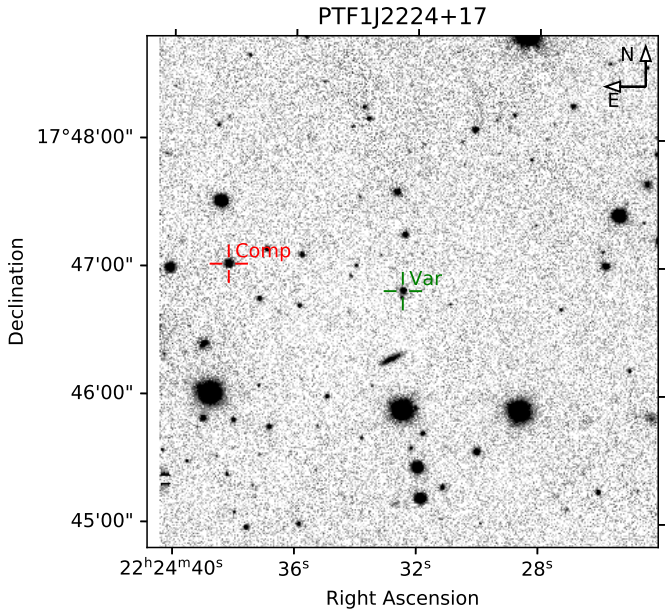}}
\caption{Finding chart identifying Ptf22 at coordinates 22:24:32.48, +17:46:48.8 (J2000) and the photometric comparison star. The size of the PanSTARRs $g$-image is $4\times4$\,arcmin.
 \label{f:chart}}
\end{figure}

While non-magnetic CVs were historically discovered through their optical variability and magnetic ones very often through their X-ray emission \citep[see e.g.][for past and recent examples]{beuermann+schwope94,falanga+19,webb+18}, contemporaneous optical transient surveys like the CRTS, the PTF, and the ZTF lead to the discovery of all kinds of CVs. 

In follow-up of distinctive variable objects found in the PTF \cite[Palomar Transient Factory,][]{law+09}, \cite{margon+14} discovered six previously undetected CVs or CV candidates, that they classify as likely polars based on strong HeII emission in their identification spectra and the likely periodicities in the range between 0.87 to 3.9 hours, interpreted as orbital period.

One of their candidates, PTF1J2224$+$17 (hereafter Ptf22) attracted our immediate interest due to a pronounced cyclotron line in its identification spectrum, which immediately revealed its classification as a polar. The tentative period of $0.87^h$, derived from a period analysis of the PTF data, however, does not fit into the picture. Such a short period cannot be orbital. In polars the orbital and the spin period agree within a few per cent. One possibility to explain the unusual observation would be that the observed signal is the spin period of the white dwarf in a intermediate polar (IP, a magnetic CV with non-synchronized white dwarf). It would be the first to show cyclotron lines and allow the first direct detection of the magnetic field in an IP. The alternative possibility would be that the object indeed is a polar. If true, its period should be $>80$\,min. This could imply that the reported periodicity is a statistical fluke.

We thus decided to perform time-resolved photometry with the aim to uniquely determine the period and to safely classify the object.

\begin{figure}[t]
\centerline{\includegraphics[width=0.55\textwidth,clip=]{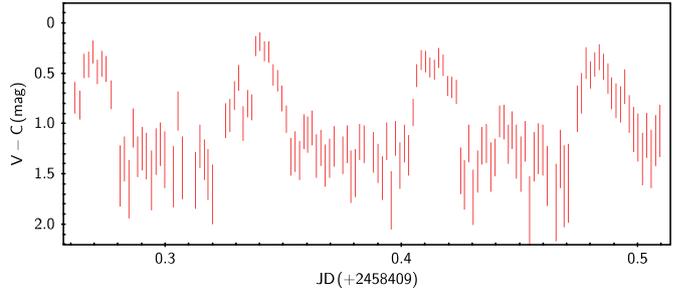}}
\caption{Differential photometry of Ptf22 obtained on October 17, 2018. \label{f:lcoct17}}
\end{figure}

\begin{table}
\caption{Photometric observations of Ptf22 from Inastars observatory\label{t:obs}}
\begin{tabular}{lcrr}
\hline
Date & Observation interval & \# frames & $T_{\rm exp}$ \\
yyyymmdd & (JD start - end) & & [s]\\
\hline
20181017 & $ 2458409.26178 - .50882$ & 126 & 150\\
20181019 & $ 2458411.24474 - .43744$ & 102 & 150\\
20181022 & $ 2458414.21424 - .31924$ & 69 & 120\\
20190102 & $ 2458486.27474 - .39184$ & 66 & 150 \\ 
\hline
\end{tabular}
\end{table}
\section{Photometric observations of Ptf22}
\subsection{Phase-resolved photometry from Inastars}
Ptf22 was observed during 3 nights in October 2018 and one additional night in January 2019 with the 14 inch Celestron reflector of Schmidt Cassegrain type located at Inastars Observatory Potsdam (IAU MPC observer code B15). The telescope is permanently installed at the roof of a one-family dwelling in a suburb of Potsdam, Germany.

All observations were done in white light. An ASTRONOMIK CLS filter\footnote{https://www.astronomik.com/en/visual-filters/cls-filter.html} was inserted to block strong emission lines at this light-polluted site. Individual images of the field around the target were recorded with an SBIG ST-8XME CCD as detector. The camera was always used with a $3\times3$ binning to reduce time overheads and with integration times of 150\,s and 120\,s, respectively. The start time of each exposure and the exposure time was written into the fits headers. The computer equipment was correlated with a time signal of atomic clocks every five minutes via the Network Time Protocol. 

CCD data reduction followed standard procedures and included dark subtraction and flatfield correction. The analysis of the light curves, i.e.differential photometry with respect to the comparison star SDSS\,J222438.20$+$174701.1, listed in the photometric database of the Sloan Digital Sky Survey with ugriz magnitudes of 19.18, 18.22, 17.56, 16.99, 16.73, respectively, was performed with C-MUNIPACK \footnote{MotlD., http://c-munipack.sourceforge.net/}. 

A summary of the observations reported in this paper is given in Table~\ref{t:obs}, which lists the observation interval per night, the JD covered, the number of frames obtained, and the used exposure times. A chart identifying the variable object and the comparison star is reproduced in Fig.~\ref{f:chart}.

\begin{figure*}[t]
\centerline{\includegraphics[width=0.66\textwidth,angle=-90]{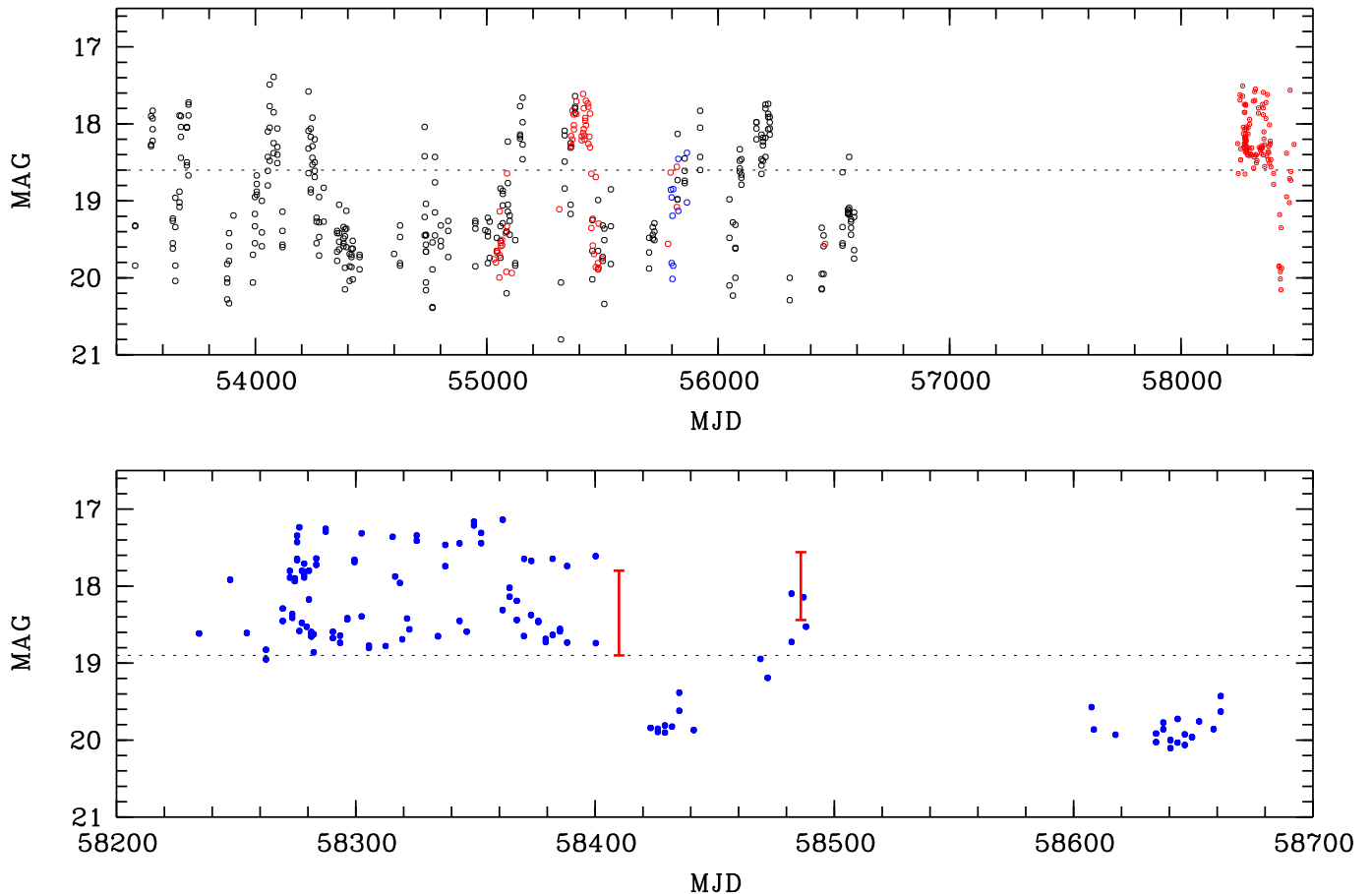}}
\caption{Archival photometry of Ptf2224. The upper panel shows white-light CRTS data (black), $R$- and $g$-band photometry from the PTF (red and blue circles) and $r$-band photometry from the ZTF (red lozenge). In the lower panel a zoomed version is shown which contains only the ZTF $g$-band data. In that panel the vertical red bars at MJD 58410 and 58486 indicate dedicated phase-resolved photometry obtained by us. Dotted horizontal lines in both panels indicate the brightness threshold separating  high and low states.\label{f:lcs}}
\end{figure*}

\subsection{Ptf22 in transient surveys}
The field of Ptf22 was covered by the CRTS \cite[326 epochs,][]{drake+09}, the PTF (62 $r$, 11 $g$), and the ZTF \cite[Zwicky Transient Facility,][]{masci+19}. We retrieved all data from the respected archives and show them in Fig.~\ref{f:lcs}. By far the largest body of data is provided by the CRTS. CRTS and PTF data show variability by 3 magnitudes which could indicate a stable periodicity. An AOV (analysis of variance) period search with the MIDAS TSA package \citep[Time Series Analysis,][]{schwarzenberg-czerny89} among the CRTS data does not reveal any significant  periodicity. The ZTF gives a clue why this happened. It shows Ptf22 to vary between $g=18.9$ and 17.3 between MJD 58230 and 58408 but then a sudden drop of brightness happened by one magnitude with a minimum around $g=20.1$. Such intensity changes are a typical property of polars, which occasionally show switches between high and low states. Past observations, those prior to MJD 56600, showed frequent changes between high and low states but the object mostly in a low state. Most of the time when Ptf22 was observed with the CRTS and a sigificant fraction of time when observed by PTF the object was in its faint low state. Mixing high and low state data clearly hampers a periodogram analysis. 

\begin{figure}[t]
\centerline{\includegraphics[width=0.23\textwidth,angle=-90,clip=]{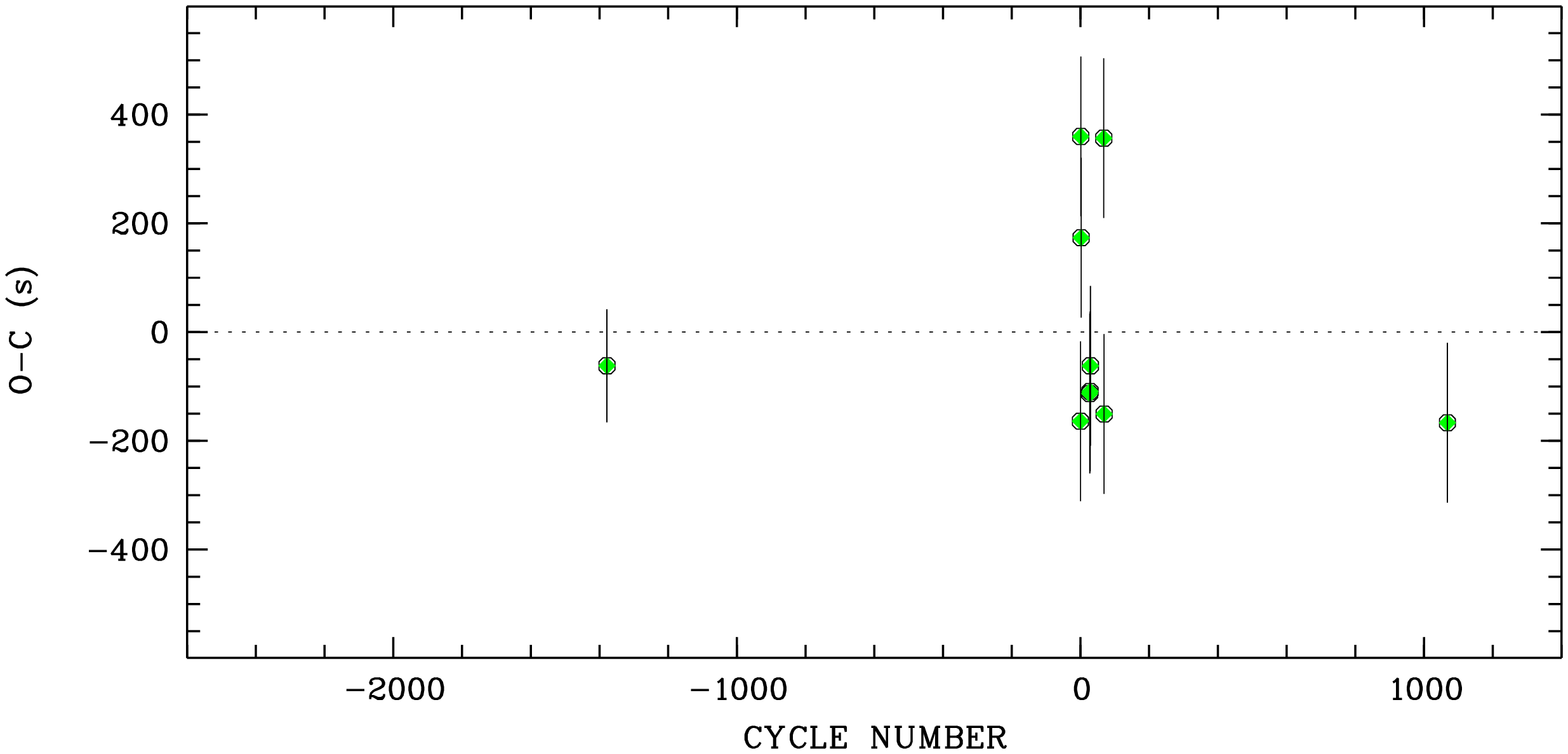}}
\centerline{\includegraphics[width=0.23\textwidth,angle=-90,clip=]{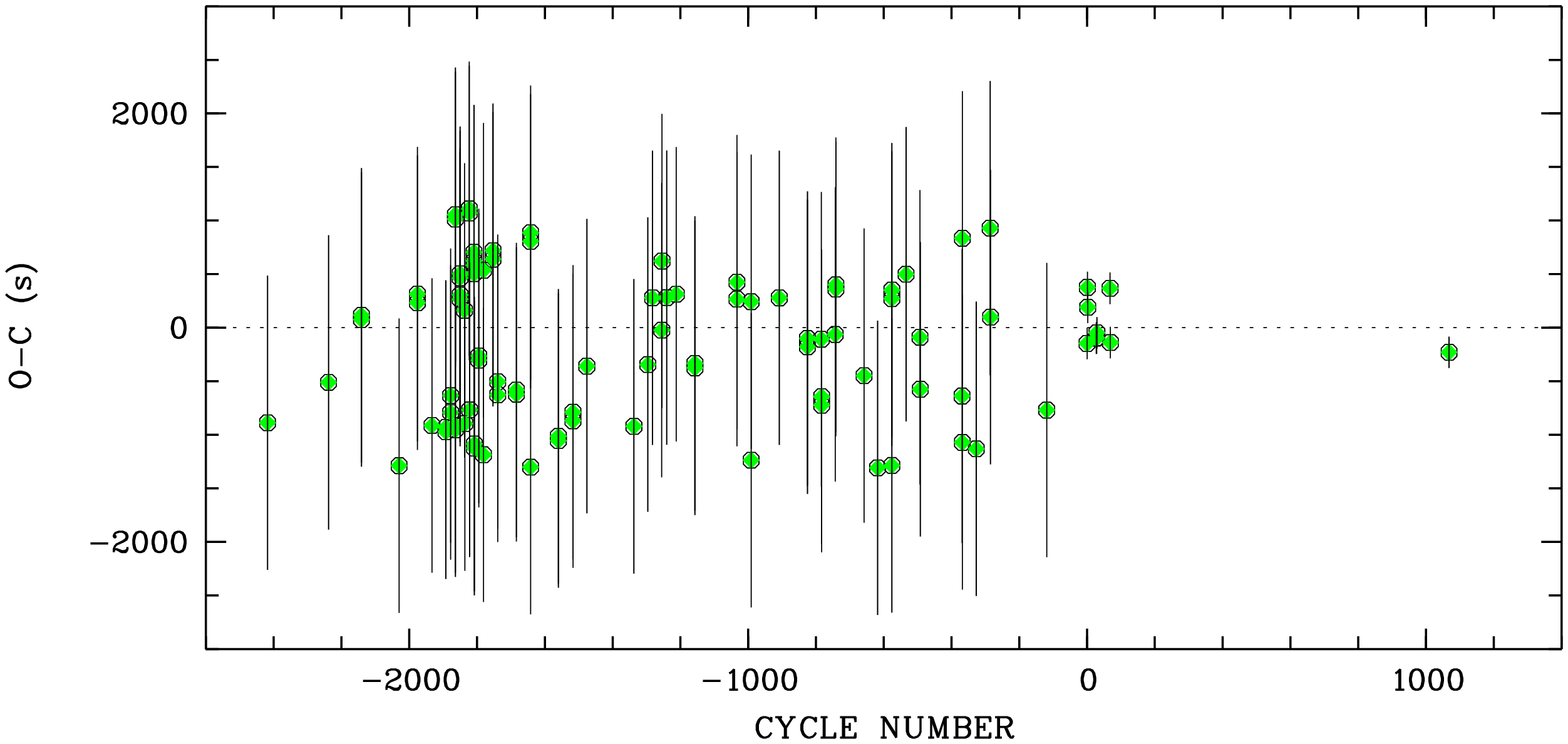}}
\caption{Observed minus calculated times of start bright phase. In the upper residuals are shown with respect to Eq.~\ref{e:eph} which uses just one, the average central, ZTF data point based on a preliminary ephemeris. In the lower panel the residuals are shown with respect to Eq.~\ref{e:ephi}. Note the different scales along the y-axis of both panels.\label{f:omc}}
\end{figure}

\section{Period analysis}

\begin{figure}[h]
\centerline{\includegraphics[width=0.2\textwidth,angle=-90,clip=]{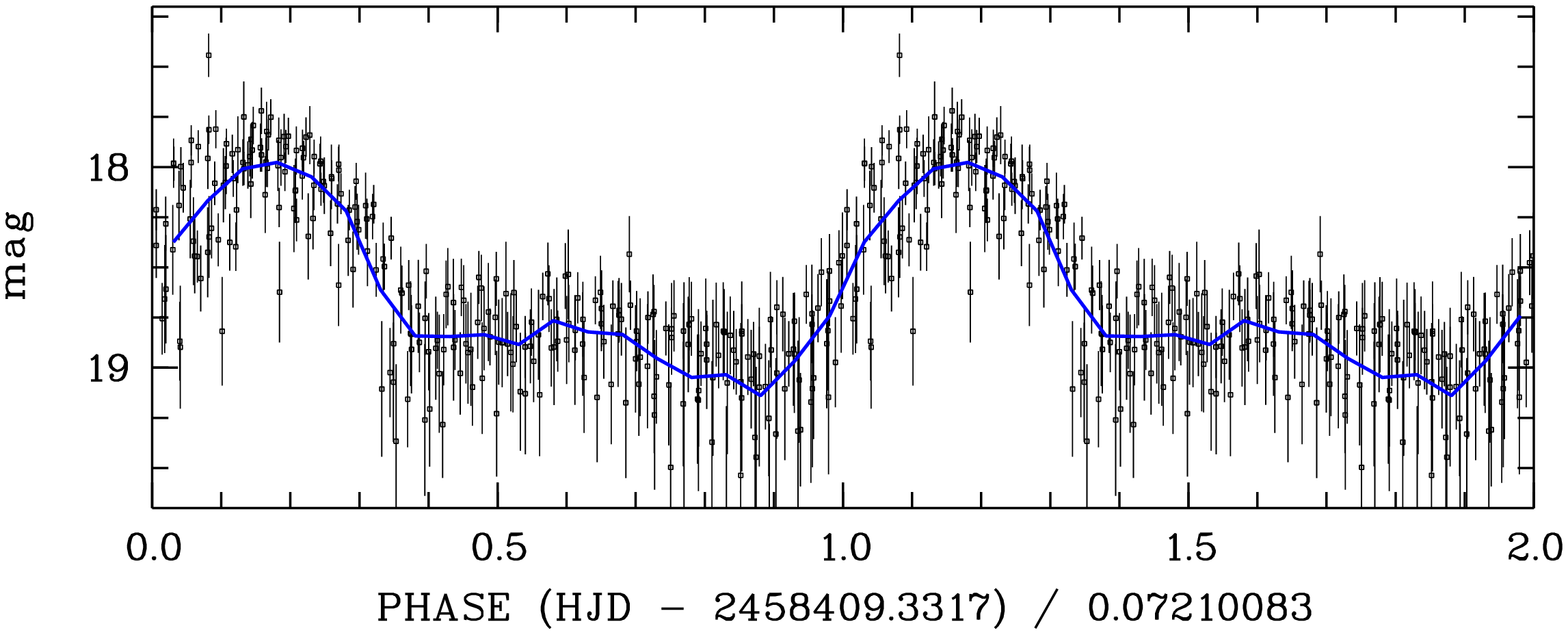}}
\centerline{\includegraphics[width=0.5\textwidth,angle=00,clip=]{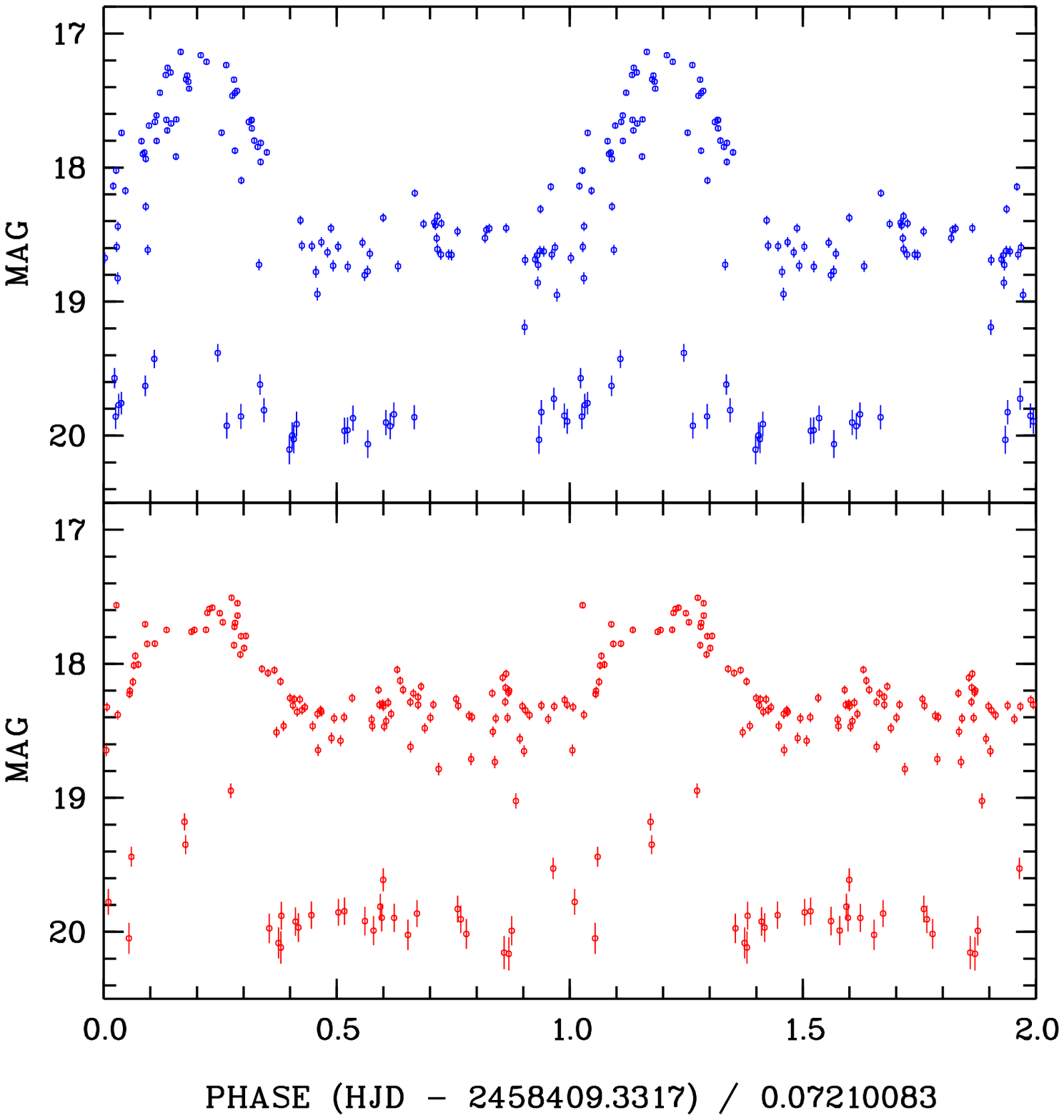}}
\caption{Folded light curves obtained from Inastars observatory obtained in 2018 (top) and the ZTF (bottom). 
In the upper panel a binned light curve with 20 phase bins is also shown. Blue and red symbols in the two lower panels indicate observations through the $g$- and $r$-filters, respectively. The mid-exposure times of original individual data were converted to phase using the ephemeris given in Eq.~\ref{e:ephi}
\label{f:lcfol}. }
\end{figure}

Already the first night of observations (Oct 17, 2018) revealed a clear repetitive pattern of the brightness change with a periodicity of about 103\,min. The light curve displayed in Fig.~\ref{f:lcoct17} has a faint phase lasting a bit less than 70\%\ of a cycle with little variability and a bright phase hump that has a slightly different shape from cycle to cycle. The brightness increase at the beginning of the bright phase seems to be steeper than the decrease towards the faint phase. 

Further observations of the object were done in following nights confirming the stability of the periodic pattern. An AOV (analysis of variance) period search using Peranso \citep{paunzen_vanmunster16} revealed the unique period of 103.4\,m. 

Among the ZTF-data we then selected only the high state data, i.e.~data with mag$<18$, and also performed a period analysis with the MIDAS TSA package \citep[Time Series Analysis,][]{schwarzenberg-czerny89} to find $P = 103.82$\,min for both the data obtained through the $g$ and $r$ filters, respectively, consistent with our dedicated phase-resolved photometry. 

\begin{table}
\caption{Start bright phase in Ptf22\label{t:times}. The first data point is an average ZTF data point, all other points are from Inastars.}
\begin{tabular}{rlr}
Cycle & $T$ (HJD) & $\Delta T$ \\
\hline
$-$1378 & 2458309.9766 & 0.0012 \\
     0  & 2458409.3292 & 0.0017\\
     1  & 2458409.4074 & 0.0017\\
     2  & 2458409.4773 & 0.0017\\
    27  & 2458411.2765 & 0.0017\\
    28  & 2458411.3486 & 0.0017\\
    29  & 2458411.4213 & 0.0017\\
    68  & 2458414.2380 & 0.0017\\
    69  & 2458414.3042 & 0.0017\\
  1068  & 2458486.3319 & 0.0017\\
\hline
\end{tabular}
\end{table}

To obtain a unique timing solution connecting all available data Inastars and ZTF-data we proceeded as follows. The start of the bright phase was defined as fiducial mark in the Inastars light curves. The very first rise to the bright phase observed on October 2017 was defined as zero point of the cycle count. The start times of all bright phases observed by us were determined individually with a cursor on a graphics display. We estimated the uncertainty to be given by the time resolution of our data, i.e.~150\,s. A linear least-squares fit for the October 2018 data was of sufficient precision to determine the cycle number of the data obtained in January 2019 without cycle count error. All those individual times derived and their finally used cycle numbers are listed in Table.~\ref{t:times}. A second linear least-squares fit including the January 2019 data point was of sufficient accuracy to determine the binary cycle of each ZTF data point without cycle count alias.

To include the ZTF-data we followed to ways. In the first, only one additional time was determined for the start of the bright phase in phase-folded, high state ($<18$ mag) ZTF-data using the preliminary ephemeris from Inastars. We used as reference time the center of the ZTF data train which corresponds to cycle $-1378$. We estimate the  timing uncertainty of this one additional data point to be 100\,s. 

A linear regression using all data points listed in Table~\ref{t:times} then yields the linear ephemeris for the start bright phase
\begin{equation}
\label{e:eph}
\mbox{HJD} = 2458409.3311(5) + E \times 0.07209997(78)
\end{equation}
where the numbers in parenthesis gives the uncertainties in the last digits. Time zero refers to the first fully covered bright phase obtained from Inastars on October 17, 2018. The residuals with respect to this linear fit are shown in the upper panel of Fig.~\ref{f:omc}.

A second attempt made use of all the bright-phase, high-state data obtained from the ZTF individually. From the preliminary phase-folded ZTF light curve all data points within the bright phase and above magnitude 18 were chosen. Each of these times can be regarded as start of the bright phase after applying a certain correction. Their times were corrected by half the length of the bright phase with an uncertainty of the same size (24\,min). This procedure revealed 86 additional data points which were used for a linear regression among a total of 95 data points to reveal the ephemeris for the start of the bright phase as 
\begin{equation}
\label{e:ephi}
\mbox{HJD} = 2458409.3317(5) + E \times 0.07210083(93)
\end{equation}
The linear fit has a reduced $\chi^2_\nu$ of 0.46 for 93 d.o.f.~indicating that the estimated errors for the additional 86 data points were too large. However, both results agree within their errors. 

We used the ephemeris given in Eq.~\ref{e:ephi} to create phase-folded light curves for the data obtained from Inastars observatory and from the ZTF that are shown in Fig.~\ref{f:lcfol}. To convert differential white-light magnitudes from Inastars to apparent magnitudes we used the SDSS $r$-band magnitude of the comparison and did not correct for any color term. In the figure all data from ZTF are shown and include high and low states.
Both data sets show that the high-state bright phase is not symmetric but saw-tooth shaped with a steeper increase and a more moderate decrease. The amplitude of the bright phase is about 1.5 mag with a maximum at $g\sim17.2$ and $r\sim17.6$. 

All the data indicate some additional variability during the faint phase which might indicate emission originating from a second pole. The data obtained in the low state at around $20^{\rm th}$ magnitude still show some enhanced brightness between phase 0 and phase 0.4 which gives evidence for some residual emission during a low state at the same accretion region.

\begin{figure}[t]
\centerline{\includegraphics[width=0.17\textwidth,angle=-90,clip=]{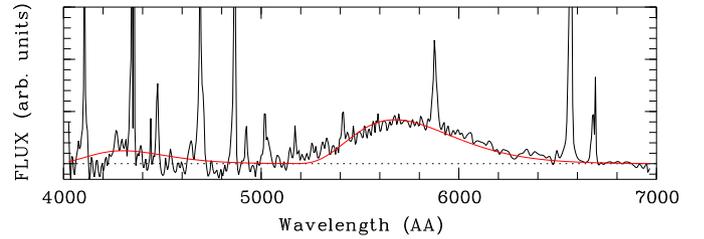}}
\caption{Cyclotron model fit for Ptf22 assuming $kT =8$\,keV and a viewing angle of 70\degr. \label{f:cyc}}
\end{figure}

\section{Spectral analysis}
The identification spectrum of \cite{margon+14} displays one prominent cyclotron hump at 5800\,\AA\ and a less prominent one at 4300\,\AA\ superposed on a blue continuum. The spectrum covers the wavelength range 4020 -- 6970\,\AA. We used the published spectrum, approximated the continuum with a second order polynomial and fitted the residuals with homogeneous, isothermal cyclotron models \citep{thompson+cawthorne87} omitting the strong emission lines of hydrogen and helium. Given the restricted coverage of the cyclotron spectrum we fixed the temperature (fit attempts were made between 5 and 10 keV) and the viewing angle at $70\degr$\footnote{The viewing angle is the angle between the line of sight and the magnetic field line(s)}. We found a satisfactory fit for $kT=8$\,keV which revealed $B=65$\,MG (Fig.~\ref{f:cyc}). For this field strength the hump at 5800\,\AA\ is the 3$^{\rm rd}$ cyclotron harmonic irrespective of the chosen temperature. We find a less convincing fit if the hump is identified with the next higher, the fourth, harmonic which would give $B=48$\,MG. Be it the former or the latter, the measured field strength is at the high side of the distribution of field strength in CVs \citep[see the recent compilation in ][their figure 20]{ferrario+20}.

\section{Summary and conclusion}
Triggered by the report of an unusually short periodicity in an obvious strongly magnetic cataclysmic variable by \citet{margon+14} we obtained phase-resolved photometry during four nights in 2018 and 2019 with the aim to uniquely determine the underlying period. The phase-resolved data obtained from Inastars observatory combined with randomly sampled archival ZTF photometric data revealed a period of 103.8239(11) (Eq.~\ref{e:eph}) or 103.8252(13) (Eq.~\ref{e:ephi}) minutes, both compatible with each other at the 1$\sigma$ level. The object may safely be classified as a polar or AM Herculis star. Further properties that support this classification are the frequent changes between high and low states observed with the CRTS, the PTF, and the ZTF, and the cyclotron humps in the identification spectrum presented by \citet{margon+14}. 

The phase-folded light curves from Inastars and the ZTF still show some scatter, in particular during the faint phase, but there is one main hump that can clearly be  associated with the main accretion region. The scatter during the faint phase may indicate accretion at a second fainter pole. The time when the identification spectrum was taken is not known to us, hence we could not determine its binary phase. Thus the field strength measured by us may belong to the second fainter pole.

The case of Ptf22 is a good example to show that contemporaneous transient surveys uncover interesting objects that are difficult to classify despite being observed rather frequently. The analysis of such objects, in particular attempts to find periodicities, still need dedicated photometric follow-up. For Ptf22 the problem lies in the frequent changes between high and low states that prevented to achieve a unique timing solution.

The accretion duty cycle in Ptf22 can be estimated from the CRTS data which cover the longest time interval. We simply count the number of data points brighter than 18.6 mag to find a duty cycle of about 1/3, clearly shorter than that of the prototype AM Herculis, where it was found to be $\sim$50\% \citep{hessman+00}.  

There are further archival data from the SDSS photometric database probably obtained in a low state ($ugriz=20.04, 19.89, 19.92, 19.86, 19.35$), the UV satellite GALEX with FUV$=18.42\pm0.10$ mag, and NUV$=18.75\pm0.05$ mag and the Gaia DR2 (ID 2737597207985505664, $p= 1.81 \pm0.31$, phot$\_$g$\_$meanmag = 19.06, $bp-rp = 0.35$). 
Following \cite{bailer-jones+18} the distance is $R_{\rm est} = 559^{+130}_{-90}$\,pc. 

There are no archival X-ray data available from Ptf22. 
If we use a bright-phase brightness of $g=17.4$ and assume a flux ratio $f_{\rm X}/f_{\rm opt} = 1$, a typical value found for CVs \citep{comparat+19}, we may estimate a bright-phase X-ray flux of $f_{\rm X}({\rm 0.5-2.0 keV}) \simeq 2 \times 10^{-13}$\,erg\,cm$^{-2}$\,s$^{-1}$ in a high state, easily to be detected in the eROSITA all sky survey which has just started.

\section*{Acknowledgments}

The CSS survey is funded by the National Aeronautics and Space
Administration under Grant No. NNG05GF22G issued through the Science
Mission Directorate Near-Earth Objects Observations Program.  The CRTS
survey is supported by the U.S.~National Science Foundation under
grants AST-0909182.

ZTF is supported by the National Science Foundation and a collaboration including Caltech, IPAC, the Weizmann Institute for Science, the Oskar Klein Center at Stockholm University, the University of Maryland, the University of Washington, Deutsches Elektronen-Synchrotron and Humboldt University, Los Alamos National Laboratories, the TANGO Consortium of Taiwan, the University of Wisconsin at Milwaukee, and Lawrence Berkeley National Laboratories. Operations are conducted by COO, IPAC, and UW.

This research was made possible through the use of the AAVSO Photometric All-Sky Survey (APASS), funded by the Robert Martin Ayers Sciences Fund and NSF AST-1412587.

\bibliography{ptf.bib}%
\section*{Author Biography}
\textbf{Axel Schwope} is senior researcher and head of the X-ray astronomy group at the Leibniz-Institute for Astrophysics Potsdam (AIP) and associated professor at the University of Potsdam. He obtained his PhD degree in 1991 and the State doctorate in 2001 from the Technical University of Berlin. His main scientific interests are X-ray surveys, galactic compact binaries and isolated neutron stars.

\end{document}